\documentclass[a4paper]{revtex4}
\usepackage{graphicx}
\usepackage{fancyhdr}
\usepackage{amsmath}
\usepackage{color}

\begin{document}

\title{Exploring extended Higgs sectors by radiative corrections with future precision coupling measurements\footnote{This talk is based on the collaboration~\cite{THDM KKY full} with Shinya Kanemura and Kei Yagyu.}}

%

\author{Mariko Kikuchi}
\affiliation{Department of Physics, University of Toyama, Toyama 930-8555, JAPAN}

\begin{abstract}
In non-minimal Higgs sectors, coupling constants of the discovered Higgs boson can deviate from the predictions in the Standard Model by effects of additional scalar bosons. 
The pattern of the deviations in various Higgs boson couplings largely depends on the structure of extended Higgs sectors.
Therefore, we may be able to determine the true Higgs sector by fingerprinting the predictions on the Higgs boson couplings in each model with the future precision data.
As the expected precision is extremely high, it is essentially important to evaluate the theory predictions as accurately as possible with radiative corrections in each model. 
We calculate a full set of one-loop corrections to the Higgs boson couplings in the two Higgs doublet models with a softly-broken $Z_2$ symmetry, 
where there are four types of Yukawa interactions.  
We discuss how to distinguish four types of the model by evaluating the pattern of the deviations in the Yukawa coupling constants and also the couplings with gauge bosons. 
In addition, we demonstrate how inner parameters in each type of the model can be extracted by the future precision measurements of these couplings at the high luminosity LHC and the International Linear Collider.  
 \end{abstract}

\maketitle

\thispagestyle{fancy}


\section{Introduction}
 The LHC Run-I has led the Standard Model (SM) particle content to be completed by discovering the Higgs boson ($h$) with the 125 GeV mass~\cite{Higgs atlas, Higgs cms}.
The data tell us that properties of the Higgs boson are consistent with those of the Higgs boson in the SM within the current uncertainties.

The discovery of $h$ does not directly mean that the true Higgs sector is the SM Higgs sector. 
In the SM, the Higgs sector is assumed to be the minimum form to spontaneously break the $SU(2)_L\times U(1)_Y$ gauge symmetry into $U(1)_\textrm{EM}$.
There are possibilities for extended structures of the Higgs sector with some number and representation of additional Higgs fields; 
e.g., models with the doublet field and additional singlet fields, doublet fields and/or triplet fields.
These extended Higgs sectors, like the SM Higgs sector, can also explain the current LHC data for $h$ in some portions of their parameter regions.
Therefore it is important to determine the true Higgs sector by future experiments.

Many of new physics models beyond the SM have characteristic extended Higgs sectors.
For example, by introducing additional Higgs fields, new physics phenomena such as neutrino mass, baryon asymmetry of the universe and dark matter can be explained.
There is a strong relation between the structure of the Higgs sector and the new physics theory. 
When the true Higgs sector is determined by future experiments, we can obtain an important hint for new physics.
Therefore, the study on the determination of the Higgs sector is one of the most important studies of new physics.

It has been known that indirect tests for new physics can be performed by utilizing precision measurements of electroweak observables such as the electroweak oblique parameters at the LEP/SLC experiments.
By the discovery of the Higgs boson and the measurement of their properties, the Higgs boson couplings have been added to the list of such useful observables.   
It is important to study deviations of the Higgs boson couplings from the SM predictions in various new physics models,
because measurement accuracy of these coupling constants will be improved at future experiments such as the High-Luminosity LHC (HL-LHC) and future electron-positron collider experiments.     
At the International Linear Collider (ILC), most of the Higgs couplings are expected to be measured with typically 1 $\%$ or better accuracy~\cite{TDR, white paper, snow mass}.
In order to use these future precision measurements, we should calculate various Higgs couplings as precisely as possible with radiative corrections.
Furthermore, by evaluating the pattern of deviations in the Higgs couplings in each extended Higgs sector, we may be able to distinguish extended Higgs sector.

There have already been many studies on radiative corrections to the $h$ couplings in extended Higgs sectors.
In Refs.~\cite{MSSMs,THDM KKY letter, KOSY, KKOSY}, one-loop corrections to the $h$ couplings have been studied in two Higgs doublet models (THDMs).
Radiative corrections to the $h$ couplings in the Higgs triplet model have been calculated in Refs.~\cite{AKKYs}.

In this talk, we investigate radiative corrections to the $h$ couplings in THDMs with a softly-broken $Z_2$ symmetry as an example of extended Higgs sectors.
Under the $Z_2$ symmetry, four types of Yukawa interactions can appear depending on the assignment of the $Z_2$ charge to particles.
We calculate renormalized couplings $hVV$, $hff$ and $hhh$ ($V=W$ and $Z$, and $f=c$, $t$, $b$ and $\tau$) at the one-loop level in the on-shell renormalization scheme~\cite{KOSY}.
We discuss how to distinguish four types of the THDM by evaluating the pattern of the deviations in the Yukawa coupling constants and also the couplings with gauge bosons. 
We further numerically demonstrate how inner parameters of the models, such as masses of extra Higgs bosons and mixing parameters, can be extracted when scale factors are experimentally determined with expected uncertainties at the HL-LHC and the ILC.
We take into account current experimental bounds and theoretical bounds from perturbative unitarity and vacuum stability in our analysis.

\section{Two Higgs doublet models}
\begin{table}
\begin{center}
{\renewcommand\arraystretch{1.2}
\begin{tabular}{|c|ccccccc|ccc|}\hline
&
\multicolumn{7}{c|}{Softly-broken $Z_2$ charge}
&\multicolumn{3}{c|}{Mixing $\xi$ factor}
\\\hline
&$\Phi_1$&$\Phi_2$&$Q_L$&$L_L$&
$u_R$&$d_R$&$e_R$ &$\xi_u$ &$\xi_d$ &$\xi_e$ 
\\\hline
Type-I &$+$&
$-$&$+$&$+$&
$-$&$-$&$-$
& $\cot\beta$& $\cot\beta$& $\cot\beta$
\\\hline
Type-II&$+$&
$-$&$+$&$+$&
$-$
&$+$&$+$
& $\cot\beta$& $-\tan\beta$& $-\tan\beta$
\\\hline
Type-X &$+$&
$-$&$+$&$+$&
$-$
&$-$&$+$
& $\cot\beta$& $\cot\beta$& $-\tan\beta$
\\\hline
Type-Y &$+$&
$-$&$+$&$+$&
$-$
&$+$&$-$
& $\cot\beta$& $-\tan\beta$& $\cot\beta$
\\\hline
\end{tabular}}
\caption{Charge assignment of the $Z_2$ symmetry given in Ref.~\cite{type X} and mixing factors in the Yukawa couplings.}
\label{yukawa_tab}
\end{center}
\end{table}

In the THDMs, there are two iso-spin doublet Higgs fields $\Phi_1$ and $\Phi_2$. 
$\Phi_1$ and $\Phi_2$ receive vacuum expectation values (VEVs) $v_1$ and $v_2$, respectively, where they satisfy relations $v (\simeq 246 \textrm{GeV})\equiv \sqrt{v_1^2+v_2^2} = (\sqrt{2}G_F)^{-1/2}$ and $\tan\beta \equiv v_2/v_1$.  
After the electroweak symmetry breaking,
five physical mass eigenstates (i.e., charged Higgs bosons $H^\pm$, a CP-odd Higgs boson $A$ and two CP-even Higgs bosons $h, H$) and three unphysical Nambu-Goldstone bosons $G^\pm$ and $ G^0$ appear.

We impose a softly-broken $Z_2$ symmetry in order to forbid flavor changing neutral currents at the tree level.
There can be four different types of Yukawa interaction, depending on the assignment for the charge of the $Z_2$ symmetry.
We define the $Z_2$ charges for all the particles as shown in Table I~\cite{type X, Logan Su}.

We assume CP invariance in our present analysis.
We can write the Higgs potential as 
 \begin{align}
 V&=m_1^2|\Phi_1|^2 +m_2^2|\Phi_2|^2 
   -m_3^2(\Phi_1^{\dagger}\Phi_2+\Phi_2^{\dagger}\Phi_1)
   +\frac{\lambda_1}{2}|\Phi_1|^4
   +\frac{\lambda_2}{2}|\Phi_2|^4
   +\lambda_3 |\Phi_1|^2|\Phi_2|^2 \notag\\
 &  +\lambda_4|\Phi_1^{\dagger}\Phi_2|^2
   +\frac{\lambda_5}{2}\left[(\Phi_1^{\dagger}\Phi_2)^2+(\Phi_2^{\dagger}\Phi_1)^2\right],
 \label{Higgs pote}
 \end{align}
where all parameters (namely $m_1^2$ - $m_3^2$, $\lambda_1$ - $\lambda_5$) are real. 
$m_3^2$ is the parameter which indicates the soft-breaking scale for the $Z_2$ symmetry. 
These parameters are replaced by the physical parameters; namely, the masses of $H^{\pm}, A, H$ and $h$, the two mixing angles $\alpha$ and $\beta$ which correspond to those among CP-even Higgs fields and charged (and CP-odd) Higgs fields, respectively, the VEV $v$ and the remaining parameter $M^2$, 
where $M^2\equiv \frac{m_3^2}{\sin\beta\cos\beta}$~\cite{KOSY}.

We here explain the scale factors for SM-like Higgs boson couplings, $\kappa_X = \frac{g_{hXX}^\textrm{THDM}}{g_{hXX}^\textrm{SM}}$,
where $X$ is any field interacting with $h$. 
At the tree level, the scale factors of gauge interactions ($\kappa_V$) correspond to $\sin(\beta-\alpha)$. 
We define that the limit in $\sin(\beta-\alpha)$ approaching to unity is the SM-like limit~\cite{THDM decouple}. 
The scale factors of $hff$ coupling constants can be expressed by
 $\kappa_f = \sin(\beta-\alpha) +\xi_f \cos(\beta-\alpha)$, 
where $\xi_f$ are given in Table~\ref{yukawa_tab}. 
The formulae of $\kappa_b$ and $\kappa_\tau$ are different in each type of the THDMs.
In Ref.~\cite{KTYY}, the possibility of fingerprinting the THDMs with future data at the HL-LHC and the ILC 
has been studied in detail by using the tree level analysis.
The full analysis with radiative corrections have been studied in Refs.~\cite{THDM KKY letter, THDM KKY full}.

\section{One-loop corrected Higgs boson couplings}
In this section, we give the results of our simulations analyses for extraction of inner parameters by fingerprints of scaling factors. 
By the on-shell renormalization scheme, we calculate several scale factors at the one-loop level which are defined following as, 
 \begin{eqnarray}
 \hat{\kappa}_X \equiv \frac{\hat{\Gamma}_{hXX}[p_1^2,p_2^2,q^2]_\textrm{THDM}}
                            {\hat{\Gamma}_{hXX}[p_1^2,p_2^2,q^2]_\textrm{SM}},
 \end{eqnarray}
where $\hat{\Gamma}_{hXX}[p_1^2,p_2^2,q^2]_{\textrm{SM}(\textrm{THDM})}$ are renormalized coupling constants in the SM (THDMs). 
Details of the renormalization calculations are described in Ref.\cite{THDM KKY full}.

We here give approximate formulae of one-loop corrected $\Delta \hat{\kappa}_X$ defined as $\Delta \hat{\kappa}_X \equiv \hat{\kappa}_X -1$ because it is useful for understanding our numerical results.
We here show approximate formulae of $\Delta \hat{\kappa}_V$, $\Delta \hat{\kappa}_\tau$ and $\Delta\hat{\kappa}_c$, 
 \begin{align}
 & \Delta \hat{\kappa}_V \simeq -\frac{1}{2}x^2 
 -\frac{1}{16\pi^2}\frac{1}{6}\sum_{\Phi=A, H,H^\pm} c_\Phi \frac{m_\Phi^2}{v^2}
 \left(1-\frac{M^2}{m_\Phi^2}\right)^2, \,\,
   \Delta \hat{\kappa}_{\tau(c)} - \Delta \hat{\kappa}_V\simeq 
    \xi_{e(u)} x, 
 \label{approximate}
 \end{align}
where $x \equiv \alpha -\beta + \frac{\pi}{2} $ and $c_\Phi = 2(1)$ for $\Phi =H^\pm (A$ and $H)$. 
We can see that there are loop contributions composed of $m_\Phi^2/v^2 (1-M^2)^2$ in $\Delta \hat{\kappa}_X$.
The contributions come from the wave function renormalization for the $h$ field.
The magnitude of the loop correction strongly depends on the balance of $M$ and $v$.
If $|M|$ is comparable to $v$, the contribution gives a quadratic power-like dependence as $m_\Phi^2$.
On the other hand, if $|M|$ is much larger than $v$, the loop effect in $\Delta \hat{\kappa}_V$ reduces as $1/m_\Phi^2$ according to the decoupling theorem~\cite{appelquist}. 
$\Delta \hat{\kappa}_f$ have the same loop contributions because the wave function renormalization appears in all renormalized Higgs couplings.
For $\Delta \hat{\kappa}_b$ and $\Delta \hat{\kappa}_t$, because there are non-negligible loop corrections which come from top quark loop diagrams,
their approximate formulae are not simple as that of $\Delta \hat{\kappa}_\tau$~\cite{THDM KKY full}.

We suppose that $\Delta \kappa_V, \Delta \kappa_b$ and $\Delta \kappa_\tau$ will be measured at the HL-LHC and the ILC500 with uncertainties expected in Refs.~\cite{HLLHC hXX, TDR, white paper, snow mass}.
According to Ref.~\cite{snow mass}, measurement uncertainties of $\Delta \kappa_X$ are given as $[\sigma (\kappa_V), \sigma (\kappa_b), \sigma(\kappa_\tau)] =
 [2\%, 4\%, 2\%]$ for the HL-LHC and  
 $[\sigma (\kappa_V), \sigma (\kappa_b), \sigma(\kappa_\tau)] =
 [0.4\%, 0.9\%, 1.9\%]$ for the ILC500.
We consider a benchmark set for central values of $(\Delta\kappa_V, \Delta\kappa_b, \Delta \kappa_\tau) = (-2\%, +5\%, +5\%)$ as an example.
We here assume measured $\Delta\kappa_c$ to be a negative value, so that the bench mark set values indicate that the Higgs sector seems that of the Type-II THDM. 
 We study parameter regions in which values of $\kappa_X$ are predicted around the central values within the 1-$\sigma$ error described above by scanning inner parameters $x, \tan\beta, m_\Phi (=m_{H^\pm}=m_H =m_A)$ and $M^2$ in the Type-II THDM. 
In these analyses, we take into account constraints from perturbative unitarity and vacuum stability.
Considering current experimental bounds in addition to the theoretical bounds,
we take $\tan\beta$ and $m_\Phi$ as $\tan\beta >1$ and $m_\Phi > 300 $GeV, respectively. 

We demonstrate to extract inner parameters $x$, $\tan\beta$, $\bar{m}_\Phi$, $m_\Phi$ and $\zeta$ by using future precision data at the LH-LHC and the ILC500, where we define such as
 $\zeta \equiv 1-M^2/m_\Phi^2$ and $\bar{m}_\Phi \equiv m_\Phi \zeta$.
$\zeta$ parameter indicates the magnitude of non-decouplingness of loop effects of extra Higgs bosons.
When $\zeta$ is unity, $\bar{m}_\Phi$ is equal to the mass of extra Higgs bosons $m_\Phi$.
If $\zeta \rightarrow 0$, namely if the magnitude of $M^2$ closes to that of $m_\Phi^2$, the loop corrections of extra Higgs bosons become zero.

\begin{figure}[t]
 \centering
  \includegraphics[width=5cm]{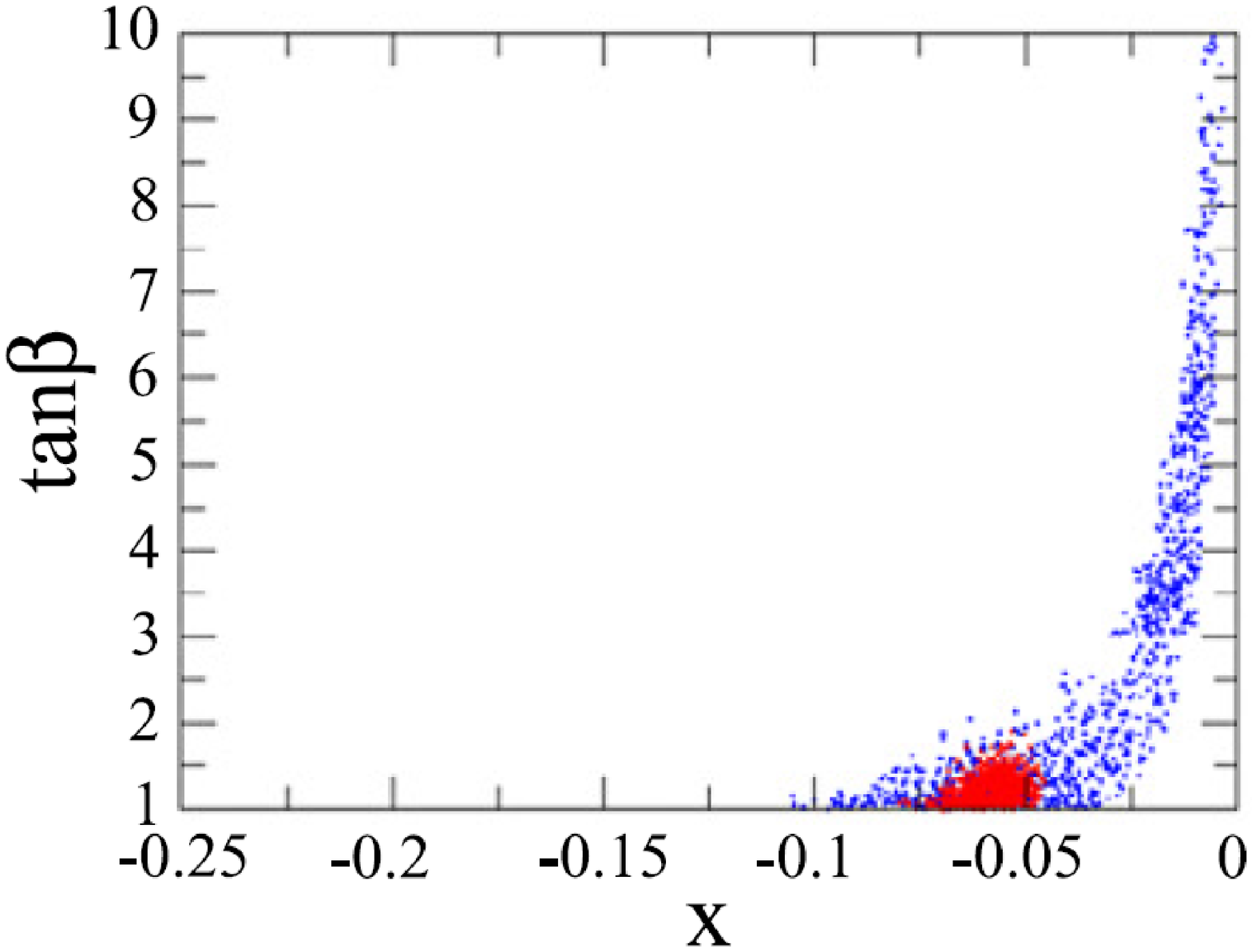}\hspace{0.3cm}
  \includegraphics[width=5cm]{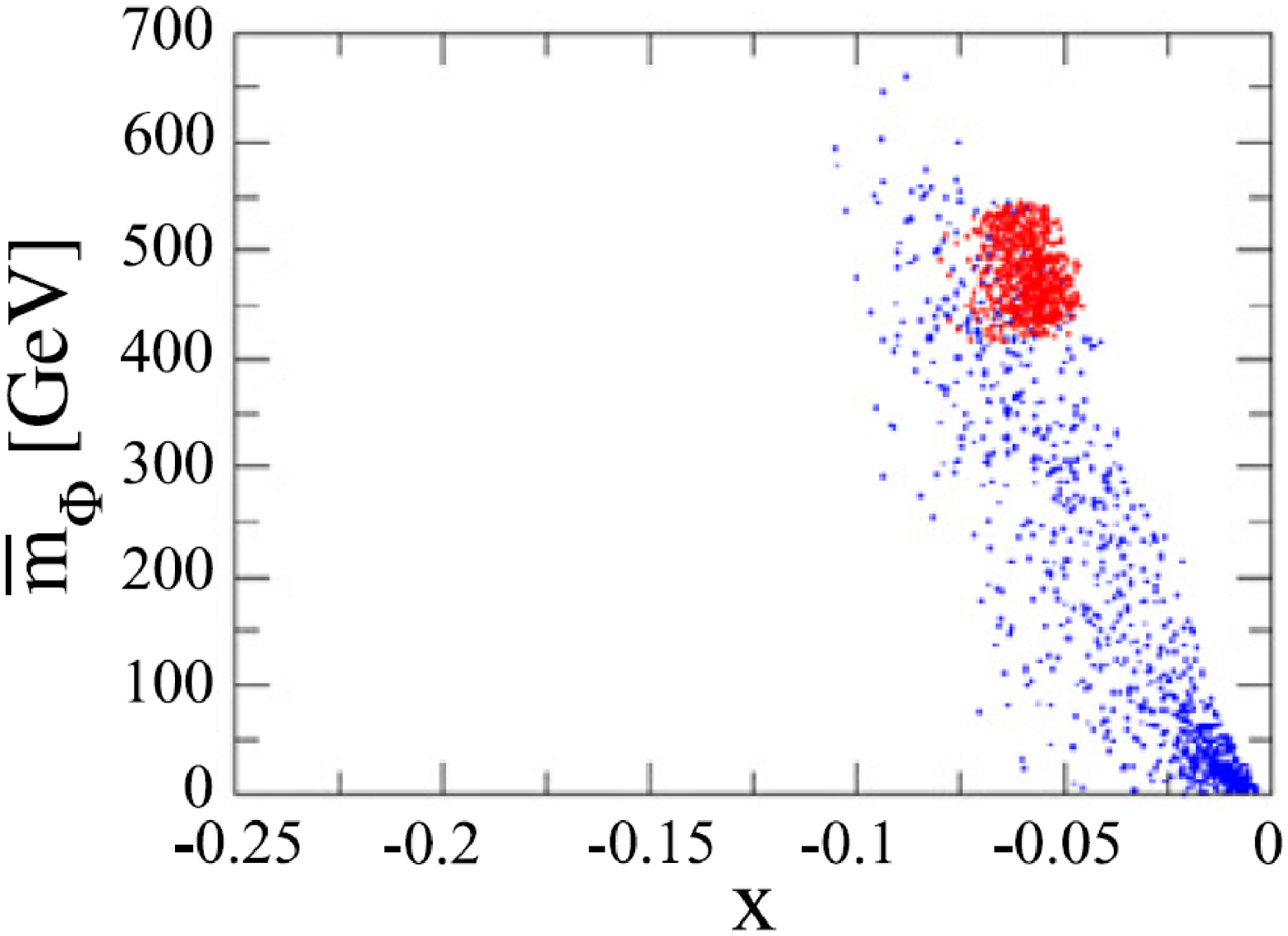}\hspace{0.3cm}
  \includegraphics[width=5cm]{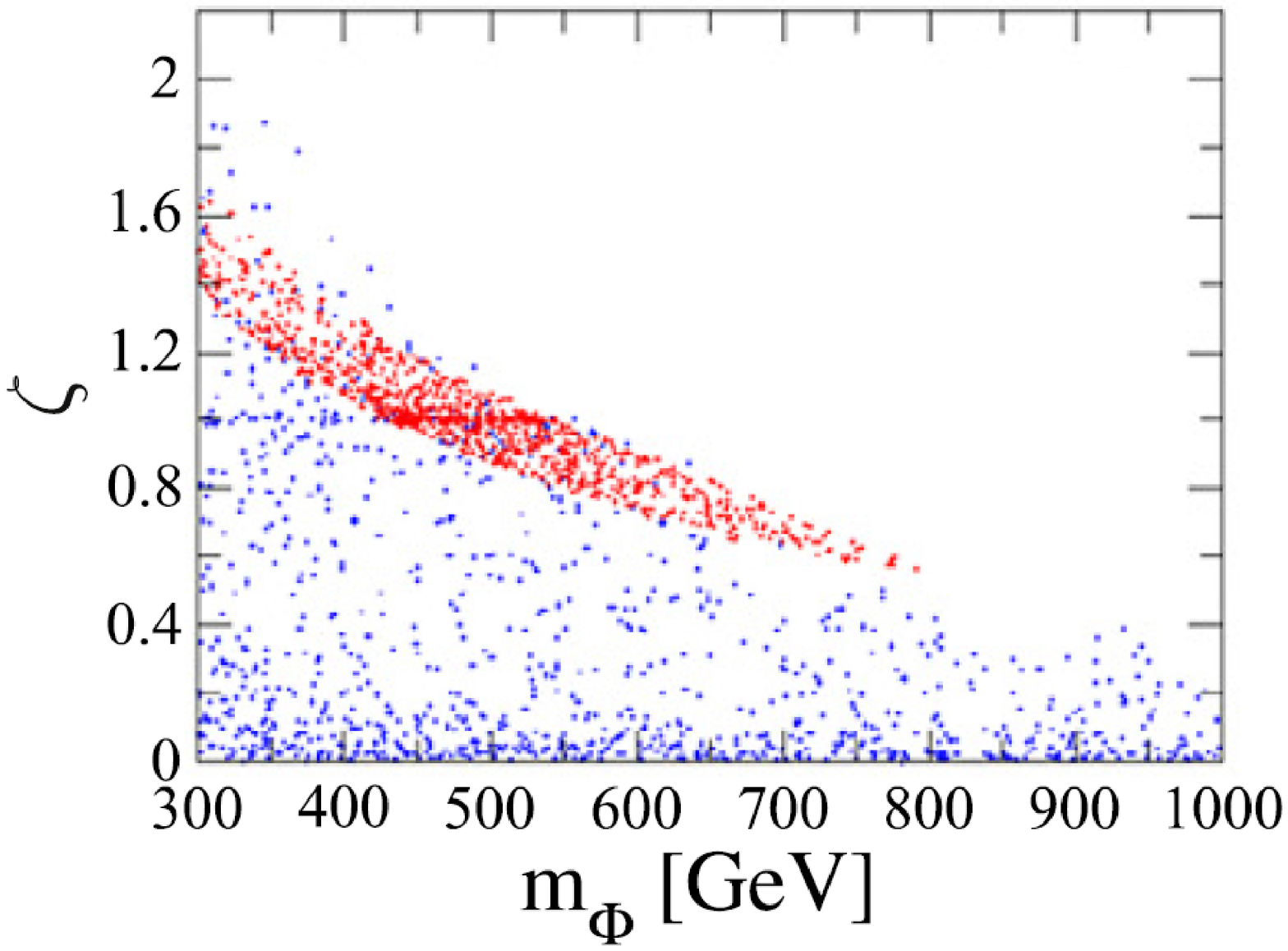}
  \caption{Scatter plots for a bench mark set $(\Delta\kappa_V, \Delta\kappa_b, \Delta \kappa_\tau) = (-2\%, +5\%, +5\%)$. Blue (red) points satisfy the bench mark set within the $1$-$\sigma$ uncertainty at the HL-LHC (ILC500) given in Ref.~\cite{snow mass}} \label{Figure}
\end{figure} 
 
Let us start to explain the left panel in Fig.~\ref{Figure}, which shows the relation between $x$ and $\tan\beta$ in the favored regions from assumed future data.
Blue points (red points) indicate arrowed regions at the HL-LHC (ILC500).
We can understand the behaviors by using the approximate formula of $\Delta \hat{\kappa}_V - \Delta \hat{\kappa}_\tau$ eq.~(\ref{approximate}).
Because values are assigned in $\Delta \hat{\kappa}_V$ and $\Delta \hat{\kappa}_\tau$, $x$ is inversely proportional to $\tan\beta$.
As you can see, points follow the inversely proportional relation.
In the case using uncertainties at the LH-LHC, the arrowed region of $\Delta \kappa_V$ is $-4\% \lesssim \Delta \kappa_V \lesssim 0\%$, which includes the SM-like limit.  
Because the value of $\Delta \kappa_V$ includes the SM-like limit, the limit $x\rightarrow 0$ is still allowed.
It means that values of mixing angles cannot be determined with precisely at the HL-LHC.
On the other hand, at the ILC500, $\Delta \kappa_V$ can be constrained to the narrow region such as $-2.4\% \lesssim  \Delta \kappa_V \lesssim -1.6\%$.
We can extract values of the mixing angles more precisely as $-0.07 \lesssim x \lesssim -0.05$ and $1 \lesssim \tan\beta \lesssim 1.8$.

The middle panel of Fig.~\ref{Figure} shows the relation between $x$ and $\bar{m}_\Phi$.
At the ILC500, regions with $x \simeq -0.06$ are allowed as described in the previous paragraph.
$\bar{m}_\Phi$ is expected to be non-zero as $400\textrm{GeV} \lesssim \bar{m}_\Phi \lesssim 550\textrm{GeV}$.
That is based on the following reasons.
$x$ and $\tan\beta$ are fixed to be $x\simeq -0.05$ and $\tan\beta \simeq 1$,
respectively, in order to satisfy $\Delta \hat{\kappa}_{\tau(b)} \simeq -5\%$.
Because it is impossible to be the non-zero value of $\Delta\hat{\kappa}_V$ by only the mixing effects, 
non-zero loop corrections are required.  
On the other hand, at the HL-LHC, $\bar{m}_\Phi$ can be still zero, because the region $\Delta \hat{\kappa}_V \rightarrow 0$ is allowed.

Finally, we discuss the behavior of the third panel in Fig.~\ref{Figure} , which shows the $m_\Phi$-$\zeta$ plane.
We can find the upper bound for the mass of extra Higgs bosons; i.e., $m_\Phi \lesssim 800$GeV at the ILC500.
There are also upper and lower bounds for $\zeta$ for each value of $m_\Phi$.
We can extract non-zero values of $\zeta$ as extraction of $\bar{m}_\Phi$ by using measurement uncertainties at the ILC500; i.e., $0.7 \lesssim \zeta \lesssim 1.6$.
Therefore, the non-decouplings effect of additional scalar bosons can be extracted. 
Such information is useful to narrow down models of new physics.  
On the other hand, at the HL-LHC, regions with $\zeta=0$ are allowed for all values of $m_\Phi$.

\section{Conclusion}
We investigated radiative corrections to the discovered Higgs boson couplings in all types of THDMs with the softly-broken $Z_2$ symmetry at the one-loop level.
We demonstrated how precisely we can extract inner parameters of extended Higgs sectors by comparing our theoretical calculations of the Higgs boson couplings with future precision measurements at the HL-LHC and the ILC500.
In our analysis, we considered theoretical constraints such as perturbative unitarity and vacuum stability as well as current experimental data.
We found that mixing parameters such as $x$ and $\tan\beta$ can be determined more precisely by using measurement uncertainties at the ILC.
Furthermore, there are possibilities to obtain the upper bound for the mass of extra Higgs bosons without their direct discoveries 
and also to get information of the decoupling property.
The determination of the Higgs sector is one of the most important studies of new physics.
In order to determine the structure of the Higgs sector by fingerprinting the Higgs boson couplings, the comprehensive study of radiative corrections to the Higgs boson couplings is an important task.

\begin{acknowledgments}
I am deeply grateful to Shinya Kanemura and Kei Yagyu for the great collaboration.
This work was supported by Grants-in-aid for JSPS,
No. 25$\cdot$10031.
\end{acknowledgments}

\bigskip 


\end{document}